\documentclass[
reprint, 
superscriptaddress, 
amsmath,
amssymb, 
aps,
floatfix,
citeautoscript
]{revtex4-2}
\usepackage{url}
\usepackage{appendix}
\usepackage{ragged2e}
\usepackage{animate}
\usepackage{graphicx}
\usepackage{xcolor}
\usepackage{adjustbox}
\usepackage[colorlinks=true,citecolor=blue,linkcolor=magenta]{hyperref}
\usepackage{bm}
\usepackage{braket}
\usepackage{lineno}
\usepackage{hyperref}

\begin{document}
\title{Quantum annealing through a first-order phase transition: field theory approach}
\author{Diego Andr\'es Rivera Orona}
\affiliation{Theoretical Division, Los Alamos National Laboratory, Los Alamos, New Mexico 87545, USA}
\author{Predrag Puno{\v{s}}evac}
\affiliation{ISR Division, Los Alamos National Laboratory, Los Alamos, New Mexico 87545, USA}
\author{Nikolai A. Sinitsyn}
\affiliation{Theoretical Division, Los Alamos National Laboratory, Los Alamos, New Mexico 87545, USA}

\begin{abstract}
\noindent

Unlike second-order phase transitions, a first-order transition has a stage, in which a system is trapped in a metastable state. The decay of this state  leads to abundant excitations over the ground state. We present a field theory for kinetics of defects emerging during quantum annealing computations. 
This theory predicts several  power laws for the error generation rate during quantum annealing either through or near the first-order phase transition. Sharp changes
between the power exponents are predicted for continuous parameter changes. Observation of this behavior would be a signature of  first-order critical points and could help identify and avoid them  for better  computations. The driven Lipkin-Meshkov-Glick model (LMGm) serves as the minimal model of interacting Ising spins that demonstrates this behavior.

\end{abstract}
\maketitle

\section{Introduction}
\label{sec:Intro}

Quantum annealing relies on the ability of a  system of qubits to follow its instantaneous ground state while the Hamiltonian is slowly varied in time. However, near a  phase transition,  the ground state must change substantially over a very short time interval. At such conditions, the system cannot remain in the  ground state, resulting in defects, such as spin qubits pointing in directions opposite to those of their ground state \cite{Zhang2024,SinitsynPokrovsky2026QuasiAdiabaticEffects}. 

The rate of defect production during a second-order phase transition is relatively well understood due to  representative solvable models \cite{SinitsynPokrovsky2026QuasiAdiabaticEffects,Polkovnikov2005,Damsky2005,Itin2009a,Polkovnikov2011,Sun2016,Suzuki2025} and the phenomenological Kibble-Zurek theory \cite{Kibble1976,Zurek2005}. They predict a power law for the number, $n_{\rm ex}$, of excitations:
\begin{equation}
n_{\rm ex}\propto N\beta^\nu,
\end{equation}
where $\beta$ is the {\it sweep rate} through the critical point of $N$ interacting qubits. Reducing $\beta$ suppresses $n_{\rm ex}$. This scaling is not dangerous, since staying in the ground state requires a time that grows only as a power law, $\tau\propto N^\alpha$.

Annealing experiments, however, show an elevated production of excitations that is captured by a  constant offset, $n_0$ \cite{Bando2020UniversalityQuantumAnnealer,King2024FrustratedIsingDynamics,Gardas2018,3bkn-v5rd}:
\begin{equation}
n_{\rm ex}(\beta)\approx n_0+c\beta^\nu.
\label{rho-scale}
\end{equation}
 Within the quantum coherent regime, $n_0$ may show only a negligibly slow logarithmic decay as $\beta\rightarrow0$. A similar defect scaling was also observed in ultracold chemical processes \cite{Sadhasivam-natcom}. The size of this contribution can be inferred by extrapolating experimentally observed power laws to zero sweep rate. How can excitations persist during very slow annealing?


\begin{figure*}[t]
\centering

\begin{minipage}[c]{0.53\textwidth}
    \centering
    \includegraphics[width=\linewidth]{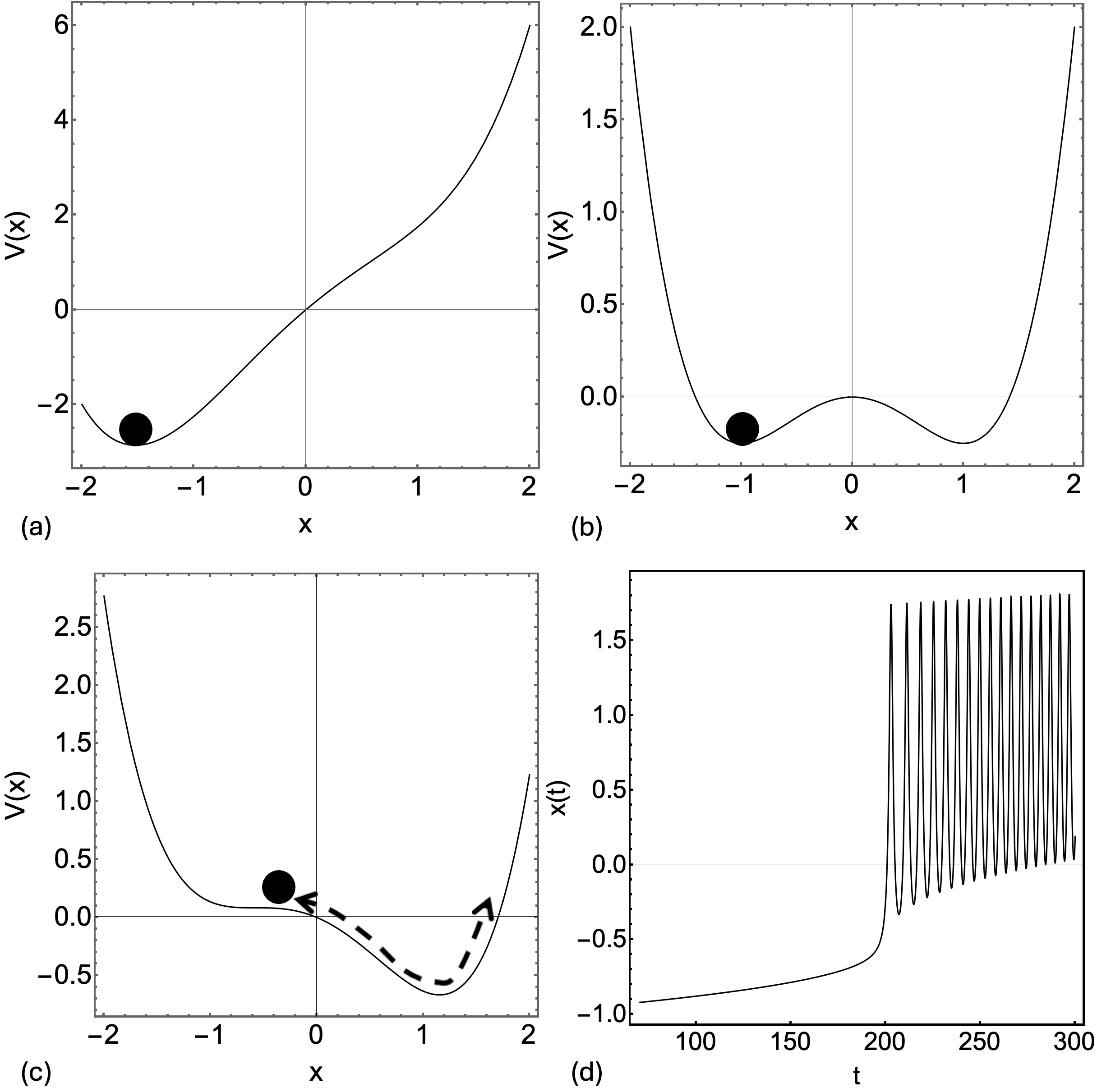}
\end{minipage}
\hfill
\begin{minipage}[c]{0.45\textwidth}
    \centering
    \includegraphics[width=\linewidth]{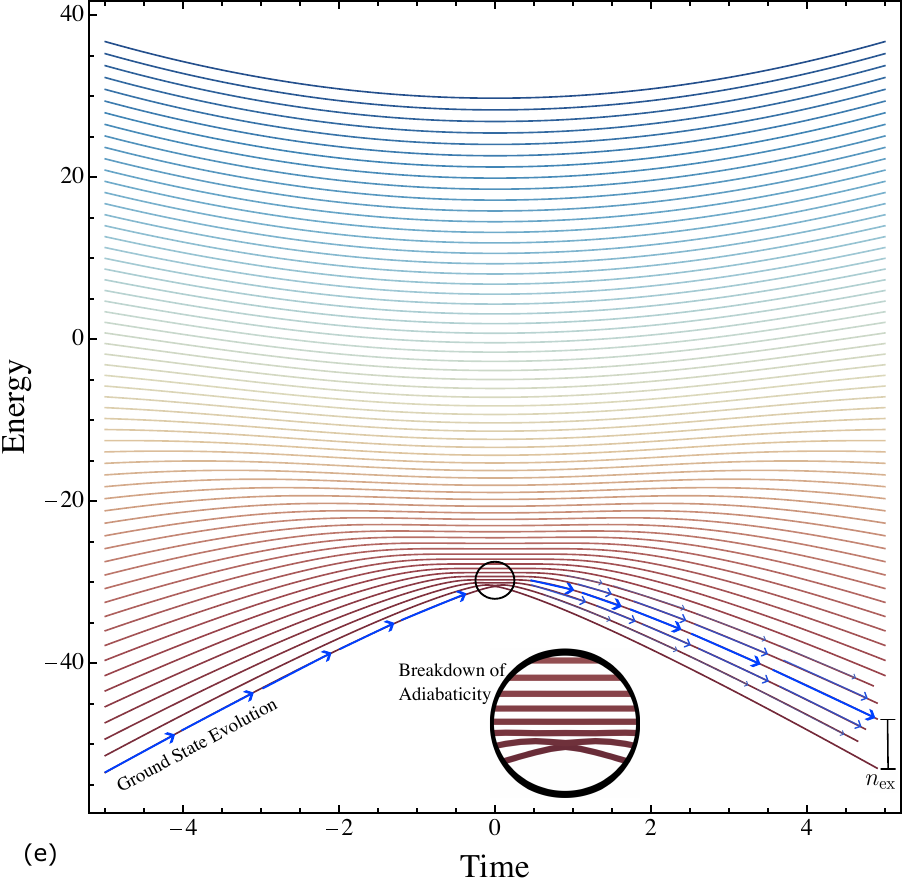}
\end{minipage}

\caption{Stages of the first-order phase transition.
(a-c) Potential energy $V(x)$ (solid curve) of the Hamiltonian, as in Eq.(\ref{hresc}), at different times: (a) For $t<0$, the system continuously follows a global minimum (black circle). (b) At $t=0$, the initial minimum of $V(x,t)$ is no longer global, but the potential barrier prevents the transition to a new global minimum. (c) At the critical moment, $t=t_c$, the initial local minimum of $V(x,t)$ disappears, and the system then falls down the potential $V(x,t)$, starting to oscillate around the global minimum. (d) A numerically computed trajectory of $x(t)$. The emergence of high-amplitude oscillations corresponds to the transition to stage (c). Here, $b=0.002$, and $x(t)$ is obtained by solving Eq.~(\ref{xt-eq}). (e) The spectrum of the Hamiltonian~(\ref{me-Hnlz}) as a function of time in the regime of a weakly first-order phase transition. Here, $S=30$, $B_x=1.0$, $D=1.1$, and $\beta=0.2$. Near $t=0$, many levels pass very close to the ground-state energy, thus enhancing nonadiabatic behavior.}
\label{stages}
\end{figure*}


First-order quantum phase transitions provide a mechanism: The system is trapped for some time in a false vacuum, i.e., a non-global energy minimum, as illustrated in Fig.~\ref{stages}(a-c). When this minimum disappears (Fig.~\ref{stages}(c)), the potential energy is released into large collective oscillations, as in Fig.~\ref{stages}(d), which later disperse into  computational errors \cite{SinitsynPokrovsky2026QuasiAdiabaticEffects}.

Thus, defects at a first-order transition are produced even in the adiabatic limit and their number is sensitive to the potential barrier height. Our field theory describes such dynamics in a broad universality class of models. 
A representative of this class is the driven Lipkin-Meshkov-Glick model (LMGm) of $N$ interacting spins \cite{Sinitsyn2024} with the Hamiltonian \cite{Liu2002,Sinitsyn2024}
\begin{equation}
H(t)=-\beta t \sum_{n=1}^N \hat{s}^z_n -\frac{D}{N} \left(\sum_{n=1}^N \hat{s}^z_n\right)^2 - B_x \sum_{n=1}^N \hat{s}^x_n,
    \label{nLZ-H}
\end{equation}
where $\hat{s}^{\alpha}_n$ are spin-1/2 operators, $D$ is the strength of Ising coupling, and $B_x$ is a transverse magnetic field.

For the LMGm, we will connect the parameters of the field theory and the quantum Hamiltonian. Figure~\ref{stages}(e) shows the time-dependent spectrum of the LMGm around the ground-energy level. Near the weakly first-order critical point, this level enters a region with a locally dense spectrum, where nonadiabatic excitations are highly probable. After leaving this critical region, the adiabaticity conditions are restored, and the system emerges on an adiabatic energy level carrying $n_{\rm ex}$ excitations above the final ground energy. Related quantum annealing models have been shown to have polynomial, exponential, and even factorial rates of energy gap closing across first-order phase transitions depending on the model itself, system size, and system parameters \cite{Joerg2010EnergyGaps, Laumann2012QuantumAdiabatic, Tsuda2013EnergyGap, Seki2012QuantumAnnealing}. 

Our analysis excludes exponentially small over-barrier tunneling effects. According to the adiabatic theorem \cite{SinitsynPokrovsky2026QuasiAdiabaticEffects}, such tunneling leads to suppression of the contribution $n_0$ at timescales $T\sim e^{fN}$, where $f>0$ is independent of $N$.  Here, our estimates are restricted to the experimentally accessible regime of $\beta/(B_xT)\ll 1$, for which quantum tunneling effects can be disregarded.

\section{Results}
\label{sec:MainResults}
When describing dynamics near a critical point, the language of quantum field theory is the most natural one \cite{SinitsynPokrovsky2026QuasiAdiabaticEffects}. Then, only variables with long-range correlations and  slow dynamics remain relevant for observable effects. For example, Ising spins with long-range interactions are described by a coarse-grained spin polarization, represented as a scalar field $\phi$ \cite{SinitsynPokrovsky2026QuasiAdiabaticEffects,CaravelliDalvit2026}. In strongly connected interaction networks, spatial dimensions become irrelevant, and this field becomes just a  variable $\phi=\phi(t)$.

Thus, the minimal model of the first-order phase transition is a real scalar $\phi^4$ theory with an external force. The Lagrangian of this field near the critical point is
\begin{equation}
L=\frac{m\dot{\phi}^2}{2}-V(\phi,t),
\label{lag}
\end{equation}
where 
\begin{equation}
V(\phi,t)=-\beta t \phi - \frac{k\phi^2}{2} + \frac{g\phi^4}{4}; \quad m,g,\beta >0,
 \label{V-lz}
\end{equation}
where $t$ is time and $\beta$ is the sweep rate of the crossing through the critical point during quantum annealing. Furthermore, $m$, $k$, and $g$ are parameters that can be derived from a more specific quantum mechanical model near the critical point, as we will illustrate in Methods for the LMGm. The first-order critical point emerges when the quadratic term in Eq.~(\ref{V-lz}) is negative, i.e., $k>0$. The case with $k<0$ is also of interest because, for negative but small $k$, the system passes in the vicinity of the critical point, enhancing nonadiabatic effects. 

This $0+1$-dimensional field theory is equivalent to one-dimensional quantum mechanics, in which $\phi$ is identified with a coordinate 
$$
x\equiv \phi
$$ of a point particle having mass $m$ and moving in the potential $V(x,t)$, as shown in Fig.~\ref{stages}(a-c).

The adiabatic limit is $\beta \rightarrow 0$. We assume that $\beta$ is small, but we are interested in nonadiabatic effects to leading order in $\beta$. The phase transition is formally encountered at $t=0$ (Fig.~\ref{stages}(b)) when the original potential minimum becomes metastable. However, the  critical point is reached later, when this minimum disappears (Fig.~\ref{stages}(c)). 
Our goal is to find $n_{\rm ex}$ long after this moment, i.e., at $t=+\infty$, assuming that, as $t\rightarrow -\infty$, the system starts in the ground state that minimizes $V(x,t)$.

As $t\rightarrow \pm \infty$, the profile of $V(x,t)$ near the ground state is well approximated by a harmonic oscillator with quantized excitations, so that the final energy is given by 
$$
E(t)=\hbar \omega(t)(n_{\rm ex}+1/2). 
$$ 
We assume that $n_{\rm ex} \rightarrow 0$ as $t\rightarrow -\infty$, and we find $n_{\rm ex}$ as $t\rightarrow +\infty$ from the amplitude of oscillations of the field after the phase transition. This number is then identified, for interacting spins, with the number of spins pointing in the direction opposite to that of their ground state.

Our main findings, derived from the field theory in Eq.~(\ref{lag}), are a set of power laws for the number of excitations.
They describe the dependence of $n_{\rm ex}$ not only on the transition rate $\beta$ but also on the characteristic distance $k$ from the boundary between different quantum annealing regimes.
These regimes correspond to 

(i) the subcritical passage at $k<0$;

(ii) touching the critical boundary at $k=0$; and

(iii) the transition through the critical point at $k>0$.

We found that  
\begin{equation}
\label{nex-km}
{\rm for}\,\, k<0: \quad n_{\rm ex} \propto \beta^{\nu_{-}} e^{-\Delta/\beta}, \quad \Delta \propto |k|^{\mu_{-}},
\end{equation}

\begin{equation}
\label{nex-k0}
{\rm for}\,\, k=0: \quad n_{\rm ex} \propto \beta^{\nu}, 
\end{equation}
and 
\begin{equation}
\label{nex-kp}
{\rm for}\,\, k>0: \quad n_{\rm ex} \propto n_0+c\beta^{\nu_+}, \quad n_0 \propto k^{\mu_+},
\end{equation}
where $c$ is some coefficient; $n_0$ does not depend on $\beta$, unless we take into account exponentially suppressed quantum over-barrier tunneling, leading to extremely slow, logarithmic in $\beta$, decay of $n_0$. The exponent $\nu_{+}$ depends weakly on the relative magnitude of $n_0$ and the $\beta$-dependent contributions in Eq.~(\ref{nex-kp}). If $n_0$ is relatively small, then $\nu_+=\nu$. In the ultra-adiabatic limit,  $\nu_+$ changes, but only a little.

The LMGm is described as the time-dependent Schr\"odinger equation with $H(t)$ from Eq.~(\ref{nLZ-H}):
\begin{equation}
i\frac{d}{dt}|\psi(t) \rangle = H(t)|\psi(t) \rangle.
\label{SE}
\end{equation}

In Methods, we show that the field Lagrangian in Eq.~(\ref{lag}) is recovered from the quantum model in Eq.~(\ref{nLZ-H}) when the number of spins is large and the coupling $D$ is close to the critical value, $D\approx B_x$, i.e., 
$$
|D-B_x|/B_x \ll 1, \quad {\rm and} \quad N\gg 1.
$$
The Lagrangian parameters in Eqs.~(\ref{lag}) and (\ref{V-lz}) are related to the parameters of the spin Hamiltonian~(\ref{nLZ-H}) as
\begin{equation}
m=\frac{1}{B_xS}, \quad k= \frac{D-B_x}{S}  , \quad g=\frac{B_x}{2S^3},
\label{dictionary}
\end{equation}
where $S=N/2$. The value $k=0$ separates the regimes with and without the first-order phase transition. According to Eq.~(\ref{dictionary}),  this critical parameter value corresponds in LMGm to the interaction strength $D=B_x$.

\begin{figure*}[t]
\centering

\begin{minipage}[c]{0.30\textwidth}
    \centering
    \includegraphics[width=\linewidth,
    trim=0.35cm 0.3cm 0.25cm 0.25cm,
    clip]{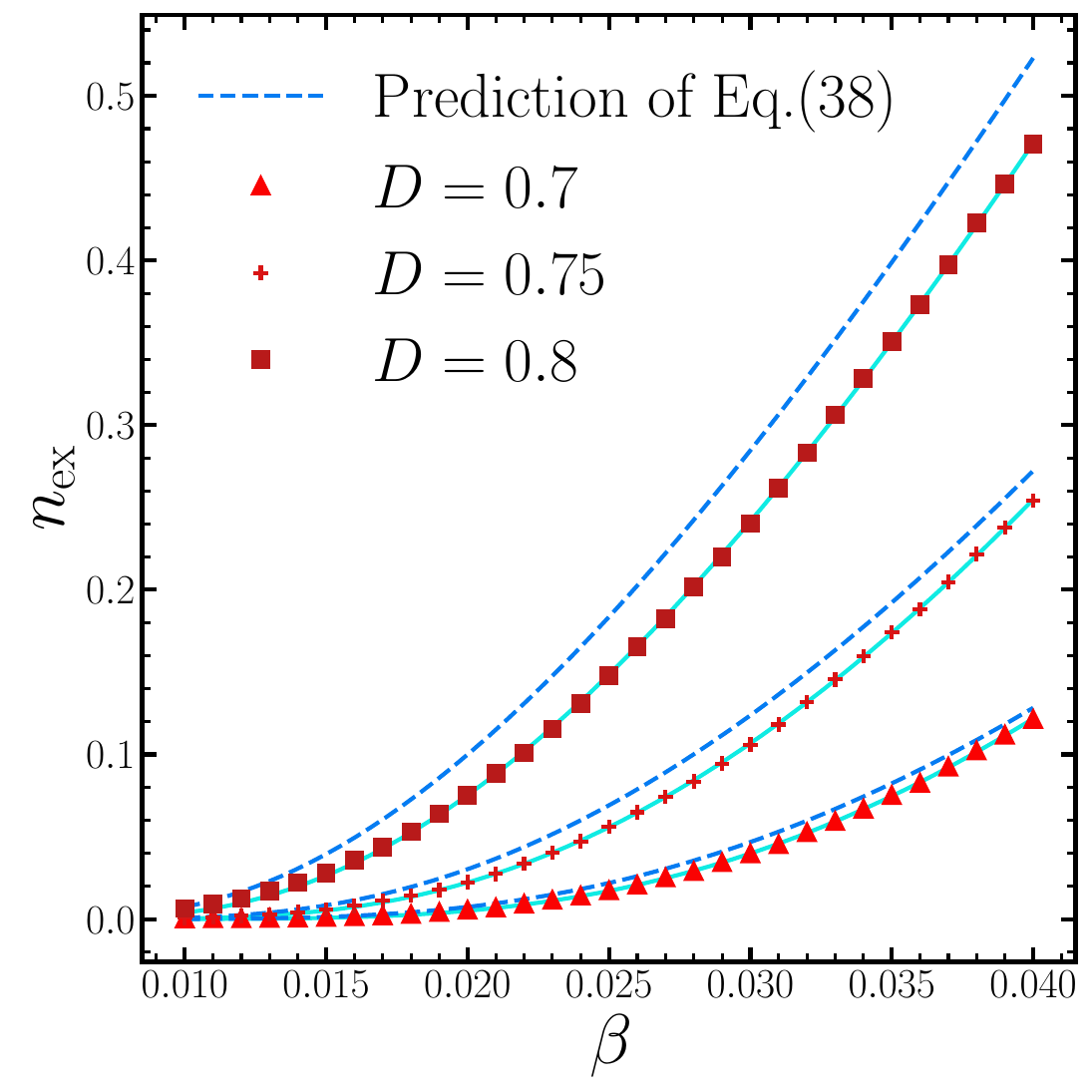}
\end{minipage}
\hfill
\begin{minipage}[c]{0.36\textwidth}
    \centering
    \includegraphics[width=\linewidth,
    trim=0.25cm 0.25cm 0.5cm 0.25cm,
    clip]{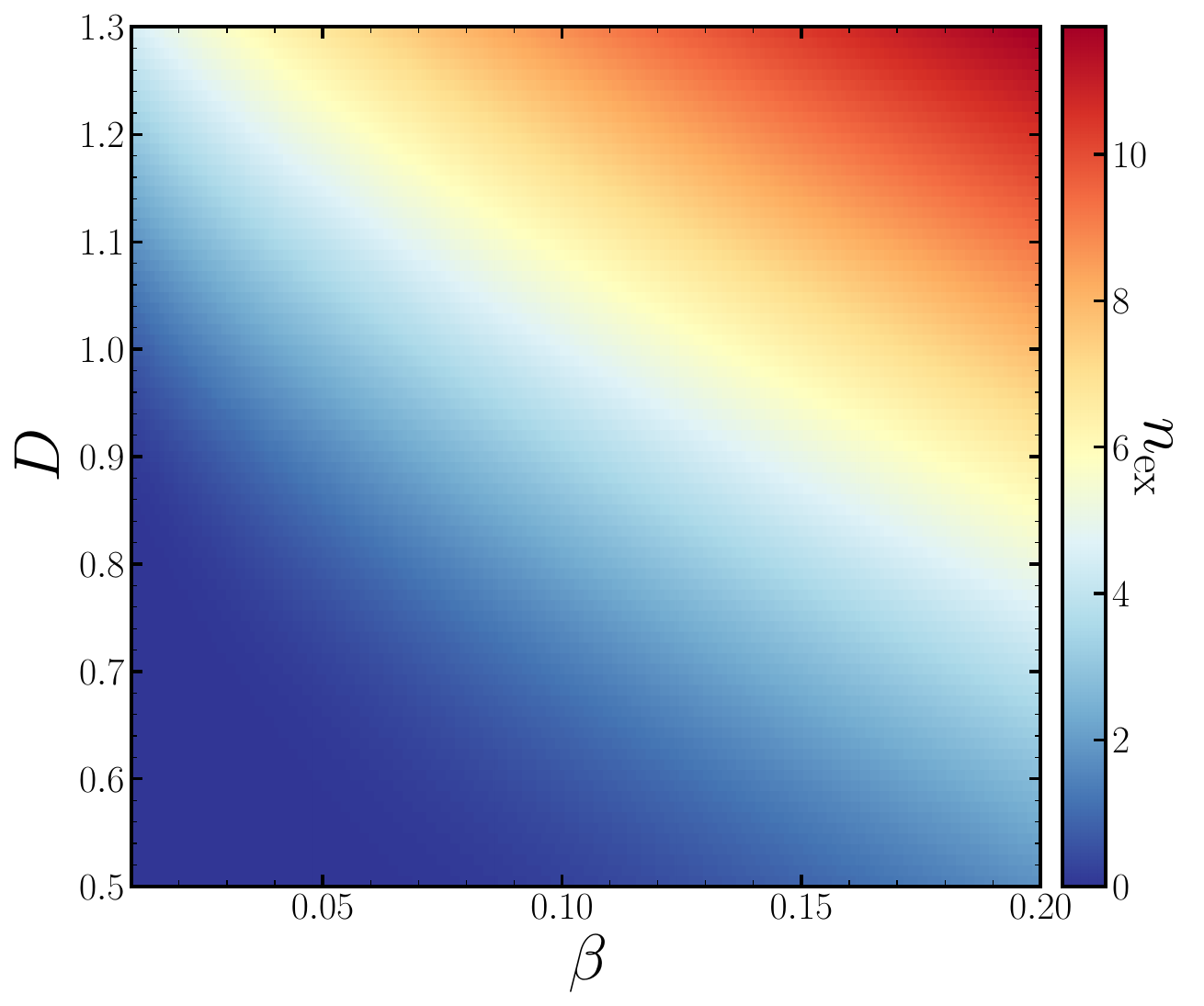}
\end{minipage}
\hfill
\begin{minipage}[c]{0.30\textwidth}
    \centering
    \includegraphics[width=\linewidth,
    trim=0.5cm 0.3cm 0.25cm 0.25cm,
    clip]{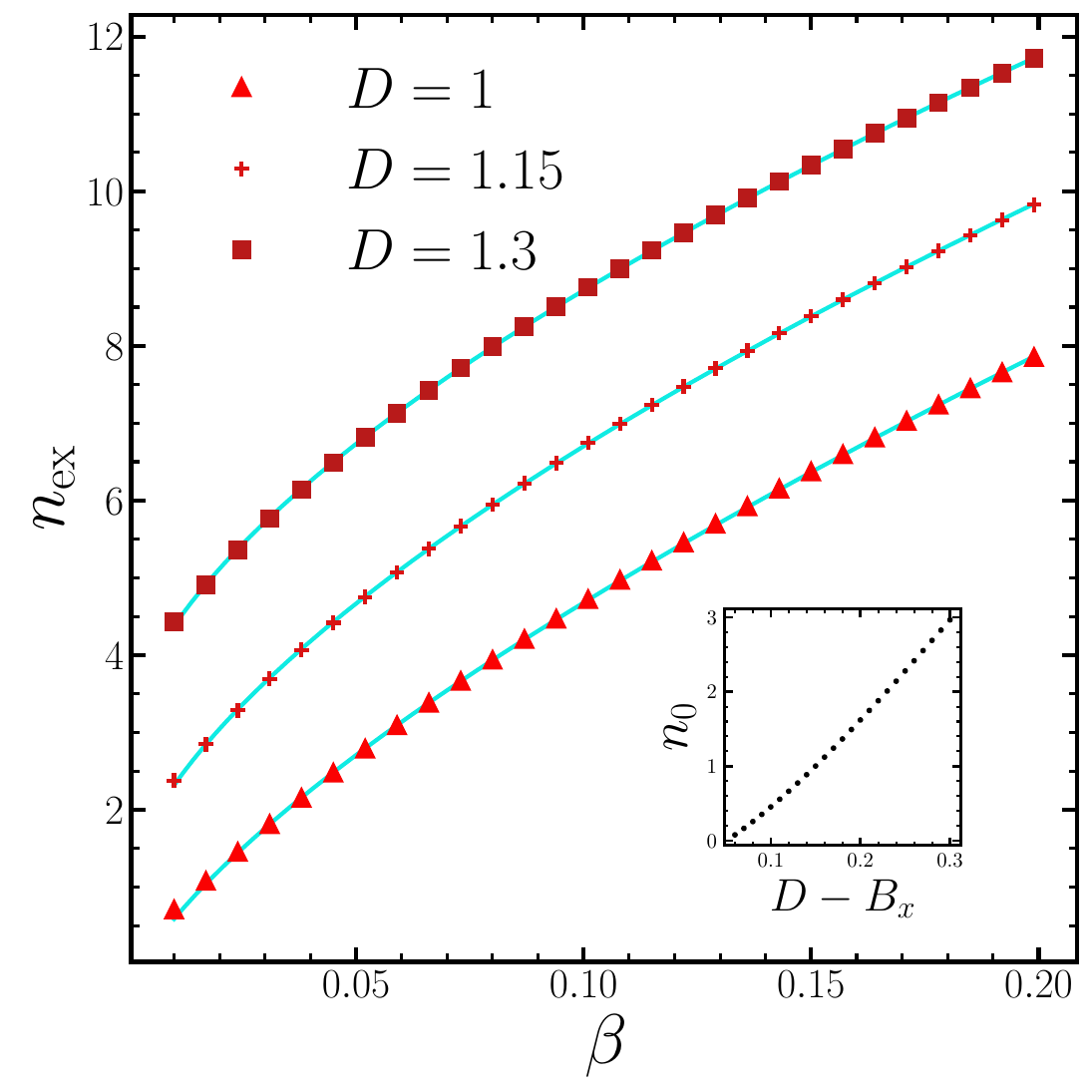}
\end{minipage}

\caption{Left Panel: The number of excitations as a function of the sweep rate for three different Ising spin couplings $D$ below the critical point. Each set of data also includes a solid, two-parameter fit line, $n_{\rm{ex}}=c\beta\exp(-\Delta/\beta)$. The dashed lines are the analytical predictions from Eq.~(\ref{eq:nex_below}) {\it without} free-parameter fitting. Center Panel: Density plot showing the number of excitations as a function of the sweep rate $\beta$ and the Ising spin coupling $D$. Right Panel: The number of excitations as a function of the sweep rate for three different Ising spin couplings. Each set of data also includes a three-parameter best-fit line, $n_{\rm{ex}}=n_0+c\beta^\nu$. The inset shows the fit values $n_0$ as a function of $D-B_{x}$. Due to finite-$N$ renormalization of the critical interaction value, $n_0=0$ is found at $D-B_x$ slightly bigger than zero.   In all simulations, we set $B_x=1$ and $S=30$.}
\label{fig:nex_vs_beta}
\end{figure*}

Figure~\ref{fig:nex_vs_beta} shows the results of a numerical simulation of the LMGm for $N=60$. The average number of excitations was found as the number of spins finally pointing in the direction opposite to that of their ground state as $t\rightarrow +\infty$. Figure~\ref{fig:nex_vs_beta} (Left) confirms that, in the subcritical regime, $n_{\rm ex}$ is suppressed exponentially as $\beta \rightarrow 0$, but there is a broad range of $\beta$ values with a power law for $n_{\rm ex}(\beta)$. Figure~\ref{fig:nex_vs_beta} (Center) is the density plot for $n_{\rm ex}$ as a function of the sweep rate $\beta$ and the spin-spin interaction parameter $D$. It shows that, as $\beta \rightarrow 0$, the critical point is found at $D=B_x=1$. For $D>B_x$, the excitation number is not visibly suppressed even in the adiabatic limit. Figure~\ref{fig:nex_vs_beta} (Right) shows the regimes of crossing the first-order critical point ($D>B_x$) or touching it at $D=B_x$. Numerical results are well fit by the power law that terminates at a constant value as $\beta \rightarrow 0$. The inset shows that this contribution to $n_{\rm ex}$ has approximately a power-law dependence on the distance, $D-B_x$, of the  interaction strength $D$ from its critical value $B_x$. 

In Methods, we derive the following values for the exponents of the model in Eq.~(\ref{lag}):
\begin{eqnarray}
\nonumber \nu_{-}&=&1,  \quad \nu=3/4, \quad \nu_+\approx 3/4,\\ 
\mu_{-}&=&2,\quad \mu_+=3/2.
\label{num-exp}
\end{eqnarray}

Our theory, in addition to the exponents from Eq.~(\ref{num-exp}), provides the values of numerical prefactors for the scaling laws. However, 
for  $N=60$, we generally did not reach values of $n_{\rm ex}\gg 1$ in the quasi-adiabatic regime numerically, so quantum over-barrier tunneling and other finite size effects distorted the results slightly, as we show in Fig.~\ref{fig:nex_vs_beta}(Left) by plotting predictions of Eq.~(\ref{eq:nex_below}) from Methods as dashed curves without any free-parameter fitting. The main finite-size effect was a small renormalization of the critical value of $D$, which is $D=B_x$ in the limit $N\rightarrow +\infty$, but which experienced up to $\sim 5$\% renormalization in our numerical results. Similarly, our numerical data were insufficient to fix the exponent $\mu_+$ with sufficiently high accuracy because of the uncertainty of the position of the critical point and other finite size effects. Within this uncertainty, however, results do not contradict our theoretical estimate.

Two additional observations can be of theoretical interest. First, the subcritical passage of the model (\ref{nLZ-H}) was studied previously by means of mean-field theory, which considered the dynamics of a single spin in a regular field found self-consistently \cite{Sinitsyn2024}. Predictions of that approach were close to, but generally different from, Eq.~(\ref{num-exp}):
$$\nu_{-}^{\rm{mean\  field}}=1, \quad \mu_{-}^{\rm{mean\  field}}=3/2, \quad \nu^{\rm{mean\  field}}=2/3.$$ 
Only the value of $\nu_{-}^{\rm{mean\  field}}$ agrees with Eq.~(\ref{num-exp}).

When the true critical exponents differ from the predictions of mean-field theory, they are called {\it anomalous}. Usually, anomalous exponents are considered specific to short-range interacting lattice models, especially in 1D and 2D spatial dimensions. What is unusual about the anomalous values of the exponents in Eq.~(\ref{num-exp}) is that they describe infinite-range interactions.

Another phenomenon that may need further theoretical attention is the {\it emergence of integrability} in the effective field theories of the passage through a critical point. In Methods, we will show that, in the vicinity of the stage in Fig.~\ref{stages}(c), the field dynamics is described by the integrable Painlev\'e-I equation, while neither the motion with the entire Lagrangian in Eq.~(\ref{lag}), nor the quantum mechanical Hamiltonian (\ref{nLZ-H}) is integrable. 

Similar observations have been made  in the theory of time-dependent second-order phase transitions. Models that were intentionally constructed to be nonintegrable have produced the integrable Painlev\'e-II equation \cite{Itin2009b,Sadhasivam-natcom} or its integrable generalizations \cite{Tyagi2025,Suzuki2025} near the phase transition. Thus, integrability is a common feature of time-dependent field theories of quantum annealing.

\section{Methods}
\label{sec:Methods}
By ``infinite-range'' we mean sufficiently long-range interactions, so that spatial dimensions become irrelevant. Such interactions are frequently found in computations by quantum annealing because the most interesting computational problems correspond to finding ground states for strongly connected graphs of spin-spin interactions. Near critical points in such systems, the field-theoretical approach should often correspond to the infinite-range models that effectively have no spatial dimension due to the direct long-range interactions between qubits. Thus, our following analytical predictions should be relevant and testable with accessible quantum annealing processors.  

\subsection{Large $N$ limit of LMGm}
The Hamiltonian of this model was defined in Eq.~(\ref{nLZ-H}). We are interested here in the effect of a sweep of the external field along the $z$-axis from large negative to large positive values, i.e., $t\in(-\infty,\infty)$. Hence, the initial ground state is ``all spins pointing down.''

Since all spins  interact equally,  the total spin polarization
$$
S_z =\sum_{n=1}^{N} s_n^z
$$
behaves as  a single spin of size 
$$
S=N/2 \gg 1,
$$
with the Hamiltonian
\begin{equation}
H=-\beta t \hat{S}_z - \frac{D}{2S}\hat{S}_z^2 -B_x\hat{S}_x,
\label{large-spin-nLZ}
\end{equation}
where $\hat{S}_z$ and $\hat{S}_x$ are projection operators of spin $S$ on the $z$ and $x$ axes, respectively. We will assume for simplicity that $N$ is even, so that $S$ is a large integer. 

Consider now the basis of spin projections
$$
\hat{S}_z |n\rangle = n|n\rangle, \quad n\in -S,\ldots, S. 
$$
In this basis, the matrix elements of the Hamiltonian in Eq.~(\ref{large-spin-nLZ}) are given by
\begin{widetext}
\begin{equation}
 H_{nm}=\left(-\beta t n-\frac{D}{2S} n^2 \right)\delta_{nm} -\frac{B_x}{2}\left( \sqrt{S(S+1)-n(n-1)}\delta_{n,m+1}+
 \sqrt{S(S+1)-n(n+1)}\delta_{n,m-1}\right).
    \label{me-Hnlz}
\end{equation}
Approximations apply near the critical point, where
$
S\gg n \gg 1. 
$
We can then disregard terms of order $n/S$, and say
\begin{equation}
 H_{nm}\approx -\left(\beta t n+\frac{D}{2S}n^2 \right)\delta_{nm} -B_x\frac{\sqrt{S^2-n^2}}{2}
 \left( \delta_{n,m+1}+\delta_{n,m-1} \right).
    \label{me-Hnlz-approx}
\end{equation}
\end{widetext}

\vspace{2mm}
\paragraph*{\bf Numerical Simulations.}
 We constructed the  Hamiltonian (\ref{me-Hnlz}) as a sparse tridiagonal matrix. Then we evolved the state $\psi(t)$ using the matrix exponential:
 $$
 \psi(t+\Delta t ) = e^{-iH(t) \Delta t}\psi(t),
 $$
 which can be efficiently done using the Crank-Nicolson update (given a small enough $\Delta t$) and efficient C++ linear algebra solvers:

 $$
 \left(I+ i \frac{\Delta t}{2}H\right)\psi(t + \Delta t) = \left(I- i \frac{\Delta t}{2}H\right) \psi(t).
 $$
 
 We began with a state $\ket{\psi(-t_{\rm{max}})}=\ket{-S}$, that is, ``all spins pointing down'' at an initial time of $-t_{\rm{max}}=-500/\beta$. All results in this paper were calculated using $S=30,B_x=1.0,$ and $3\times 10^6$ time steps. Having completed the evolution, we compute the number of excitations as 
\begin{equation}
    n_{\rm{ex}} = S - \braket{\psi(t_{\rm{max}})|\hat S_z |\psi(t_{\rm{max}})}.
\end{equation}
This is the average number of spins pointing down at the end of the time evolution. 

\vspace{2mm}
\paragraph*{\bf Semiclassical approximation for large $S$.}
If we now search for the solution of Eq.~(\ref{SE}) with the Hamiltonian (\ref{me-Hnlz-approx}) in the form 
$$
\ket{\psi} =\sum_{n=-S}^S a_n \ket{n}, 
$$
then the amplitudes satisfy
$$
i\dot{a}_n = -\left(\beta t n +\frac{Dn^2}{2S} \right)a_n -\frac{B_x\sqrt{S^2-n^2}}{2}\left(a_{n-1}+a_{n+1} \right).
$$
We multiply this equation by $e^{in\varphi}$ and sum over $n$ to find an equation for the function
$$
u(\varphi,t) \equiv \sum_n e^{i\varphi n} a_n(t),
$$
such that 
$$
i\partial_t u = H\left(\varphi, -i\partial_{\varphi},t \right) u,
$$
where $n$ is replaced in the Hamiltonian operator~(\ref{me-Hnlz-approx}) by $-i\partial_{\varphi}$. Since $-i\partial_{\varphi}$ and $\varphi$ satisfy the standard momentum-coordinate commutation relations, for $N\gg 1$, we can consider the Hamiltonian as that of classical variables $({\varphi,n})$ identified as, respectively, coordinate and momentum. In our case,
\begin{equation}
H_{cl}(\varphi,n) = -\beta t n -\frac{D}{2S}n^2 -B_x\sqrt{S^2-n^2} \cos \varphi.
\label{H_cl}
\end{equation}

Small excitations live near a fixed point at $\varphi=0$ and $|n|\ll S$, so we write
$$
\cos \varphi \approx 1-\frac{\varphi^2}{2},
$$
$$
\sqrt{S^2-n^2} \approx S-\frac{n^2}{2S}-\frac{n^4}{8S^3}.
$$

It is convenient  to make a canonical transformation:  
$$
\varphi \rightarrow -p, \quad n\rightarrow x,
$$
which makes $(x,p)$ a canonical pair of variables, such that
$$
\dot{x} = \partial_{p} H_{cl}, \quad \dot{p} = -\partial_{x} H_{cl},
$$
where 
\begin{equation}
H_{cl}(x, p,t) = \frac{B_xS p^2}{2} -\beta t x-  \frac{D-B_x}{2S}x^2  +\frac{B_x x^4}{8S^3}.
\label{hcl-nlz}
\end{equation}
Thus, the Hamiltonian acquires the  form of a classical particle with mass $m$ moving in a quartic potential:   
\begin{equation}
H_{cl}(x,p,t) = \frac{p^2}{2m} -\frac{kx^2}{2}+\frac{g x^4}{4} -\beta t x,
\label{hcl1}
\end{equation}
where parameters are related to the original quantum model via Eq.~(\ref{dictionary}).

\subsection{Effectively classical motion}
The derivation of Eq.~(\ref{hcl1}) is an illustration of a deeper relation between dissipationless quantum phase transitions and effective field theories, leading, in the saddle-point approximation, to classical equations of motion. 

Defect production in the quasi-adiabatic limit of such field theories is well described by the saddle-point equations \cite{SinitsynPokrovsky2026QuasiAdiabaticEffects}, which are now equivalent to canonical classical equations
\begin{equation}
 \frac{d{x}}{dt} =\frac{\partial H_{cl}}{\partial p}, \quad \frac{d p}{dt}=-\frac{\partial H_{cl}}{\partial x}, 
\label{ham-em} 
\end{equation}
where 
\begin{equation}
H_{cl}(x, p)= \frac{p^2}{2m} +V(x,t).
\label{ham-pt1}
\end{equation}

The number of nonadiabatic excitations is inferred from the behavior of the adiabatic invariant \cite{Altland2009,Itin2009b,Sadhasivam-natcom,SinitsynPokrovsky2026QuasiAdiabaticEffects}
\begin{equation}
I=\frac{1}{2\pi} \oint p\,dx,
\label{ainv-def}
\end{equation}
where the integral is over a periodic trajectory in phase space.  Our initial conditions correspond to $I_{-\infty}\equiv I(t\rightarrow -\infty) = 0$. As $t\rightarrow +\infty$, the adiabatic invariant is related to the number of  excitations over the ground state via 
$$
n_{\rm ex} = I_{+\infty}/\hbar,
$$
where $\hbar$ is Planck's constant, which in our theory was set to $\hbar =1$, so $I_{+\infty}$ is the number of  excitations.

In the equations of motion (\ref{ham-em}), we rescale variables to 
\begin{equation}
t \rightarrow  \gamma t, \quad p\rightarrow \theta p, \quad x \rightarrow \eta x, 
\label{rescale-1}
\end{equation}
with constants
\begin{equation}
\gamma=\sqrt{\frac{m}{|k|}}, \quad \theta =\frac{|k|m^{1/2}}{g^{1/2}}, \quad \eta=\frac{|k|^{1/2}}{g^{1/2}},    \label{rescaled}
\end{equation}
such that the equations remain canonical but with a new Hamiltonian that does not depend on the constant parameters $m$, $k$, and $g$, namely,
\begin{equation}
\mathcal{H}=\frac{\gamma}{\theta \eta}H_{cl}(\eta x, \theta p;\gamma t)=\frac{p^2}{2}-b tx \pm \frac{x^2}{2} + \frac{x^4}{4},
\label{hresc}
\end{equation}
where ``$-$'' and ``$+$'' correspond to $k>0$ and $k<0$, respectively,  and 
\begin{equation}
  b=\frac{\beta g^{1/2} m^{1/2}}{|k|^2}.
    \label{b-def}
\end{equation}
A weakly first-order phase transition corresponds to $0<k\ll g^{2/3}/m^{1/3}$. Note that the adiabatic limit $b\rightarrow 0$ and the limit $|k|\rightarrow 0$ in Eq.~(\ref{b-def}) do not commute. This explains the discontinuity of power-law exponents in Eqs.~(\ref{nex-km})-(\ref{nex-kp}) across $|k|=0$.

For the LMGm with parameters from Eq.~(\ref{dictionary}) we have
\begin{equation}
  b=\frac{\beta}{\sqrt{2}|B_x-D|^2}.
    \label{bnLZM}
\end{equation}
Let 
$
\mathcal{I} \equiv \frac{1}{2\pi}\oint p\,dx
$
be the adiabatic invariant for the Hamiltonian (\ref{hresc}) 
written in the rescaled variables. Changing variables did not affect the initial conditions: $\mathcal{I}\rightarrow 0$ as $t\rightarrow -\infty$. Since $\mathcal{H}$ depends
only on  $b$, the final adiabatic invariant is also a function of 
only this combination: $\mathcal{I}=\mathcal{I}(b)$. The adiabatic invariant in the original 
variables is then given by 
$$
I(\beta)=\eta \theta \mathcal{I}(b(\beta)) = \frac{|k|^{3/2}m^{1/2}}{g}
\mathcal{I}\left(\frac{\beta g^{1/2} m^{1/2}}{|k|^2} \right).
$$

Let us assume that in the rescaled variables the final adiabatic invariant scales as 
$$
\mathcal{I}_{+\infty}=\mathcal{I}_0 + cb^{\nu_+},
$$
with certain numerical constants $\mathcal{I}_0$, $c$, and $\nu_+$. Then, in the original variables of the field theory we have
\begin{equation}
    I_{+\infty}=I_0+\Delta I=\frac{\mathcal{I}_0|k|^{3/2}m^{1/2}}{g} +\frac{cm^{\frac{1+\nu_+}{2}}}{g^{\frac{2-\nu_+}{2}}}|k|^{\frac{3}{2}-2\nu_+} \beta^{\nu_+}.
    \label{I-uni}
\end{equation}
The scaling $I_0\propto |k|^{3/2}$ in Eq.~(\ref{I-uni}) reproduces the value of $\mu_{+}=3/2$ in Eq.~(\ref{num-exp}).

\subsection{Case of subcritical $k<0$ ($D<B_x$ in LMGm)}

At negative $k$, the potential always has a single minimum, so the critical point is not crossed, but, at small $|k|$, the vicinity of the critical point influences the production of defects. In rescaled variables, the potential energy for the motion of an effective particle is
\begin{equation}
V(x)=-b tx +\frac{x^2}{2} + \frac{x^4}{4}.
\label{vxkm}
\end{equation}

At any $t$, there is now a single minimum $x=x_m$ of this potential, given by $V'(x_m)=0$, leading to an equation
\begin{equation}
bt=x_m+x_m^3,
    \label{mineq}
\end{equation}
from which we  extract a relation 
\begin{equation}
\left(\frac{dx_m}{dt}\right)^{-1}=\frac{1}{b}\left( 1+3x_m^2\right).
    \label{dxdt}
\end{equation}

Our excitations are weak, so their dynamics are well described by approximating the potential as that of a harmonic oscillator:
\begin{equation}
V(x) \approx \frac{\omega^2 (x-x_m(t))^2}{2}, \quad \omega=\omega(x_m).
    \label{veff1}
\end{equation}
By taking the second derivative of $V(x)$ in Eqs.~(\ref{vxkm}) and (\ref{veff1}) at $x=x_m$, we find 
\begin{equation}
\omega^2(x_m)=1+3x_m^2.
    \label{om-def}
\end{equation}
 Since there is a one-to-one correspondence between $x_m$ and $t$, it is convenient to switch from $t$ to $x_m$. For a harmonic oscillator, the action-angle variables are related to $(x,p)$ by
\begin{equation}
x(I,\theta) =x_m+\sqrt{\frac{2I}{m\omega}}\sin \theta, \quad p(I,\theta)=\sqrt{2Im\omega}\cos \theta.
\label{action-ang}
\end{equation}

When switching from $(x,p)$ to $(\theta,I)$, the nonadiabatic correction to the Hamiltonian
arises from the time dependence of $x_m$ in the 1-form $pdx$ inside the invariant action, $S=\int  \{pdx-\mathcal{H}dt \}$:
$$
pdx-\mathcal{H}(x,p)dt\approx Id\theta+(p(I,\theta)\dot{x}_m -\omega I) dt,
$$
where we disregard the correction due to the time dependence of $\omega$ because it appears at a higher power of the small $I$. The equations of motion in $(\theta, I)$ coordinates are then given by
\begin{equation}
\frac{dI}{dt}\approx -\sqrt{2m\omega I} \sin \theta \dot{x}_m, \quad \frac{d\theta}{dt} \approx \omega,
\label{eqm}
\end{equation}
where we again disregarded a negligible correction to the expression for $\dot{\theta}$. With such approximations, it is convenient to change variables
$$
I=J^2.
$$
The formal solution of Eq.~(\ref{eqm}) at $m=1$ is given by
\begin{eqnarray}
    \theta(x) &=& \theta_0+ \int_{-\infty}^{x} \omega(x_m) \left(\frac{dx_m}{dt}\right)^{-1}\, dx_m, \\
    J(x)&=&-\int_{-\infty}^{x} \sqrt{\frac{\omega(x_m)}{2}} \sin \theta (x_m) \, dx_m,
\end{eqnarray}
where the initial conditions correspond to $J\rightarrow 0$ as $x_m \rightarrow -\infty$, and the quantum mechanical ground state in the semiclassical calculations corresponds to averaging over all possible values of the initial angle $\theta_0$. 

Thus, the average change of the adiabatic invariant  is given by 
$$
\langle I \rangle = \frac{1}{2\pi} \int_{0}^{2\pi} (J^2)_{x \rightarrow \infty} \, d\theta_0. 
$$
For $x_m$ changing from $-\infty$ to $\infty$, with Eqs.~(\ref{dxdt}) and~(\ref{om-def}), this leads  to 
\begin{eqnarray}
\nonumber \langle I\rangle &=&\frac{1}{2}\left| \int_{-\infty}^{\infty} dx_m \sqrt{\frac{\omega(x_m)}{2}}e^{i\theta(x_m)}\right|^2\\
\nonumber &=&\frac{1}{4}\left|\int_{-\infty}^{\infty} dx_m  \, \sqrt{\omega(x_m)}e^{i\int_{-\infty}^{x_m} \omega(x) \frac{dt}{dx} \,dx} \right|^2\\
\nonumber &=&\frac{1}{4}\left|\int_{-\infty}^{\infty} dx_m  \, \sqrt{\omega(x_m)} e^{\frac{i}{b}\int_{-\infty}^{x_m} (1+3x^2)^{3/2} \,dx} \right|^2.
\end{eqnarray}
Due to the smallness of  $b$, this integral can be calculated by the method of steepest descent. The extremum is reached when the expression inside the integral in the exponent is zero, that is, at
$$
1+3x_m^2=0,
$$
which is satisfied at 
$$
x_m=x_0 \equiv i/\sqrt{3}.
$$
The leading exponent of the integral (taken as absolute square) is given by 
\begin{eqnarray}
\nonumber \langle I \rangle &\propto& e^{-\frac{2}{b} {\rm Im} \int_0^{x_0} (1+3x_m^2)^{3/2} \, dx_m}, \\
\nonumber &=& e^{-\frac{2}{b}  \int_0^{1/\sqrt{3}} (1-3s^2)^{3/2} \, ds }\\
\nonumber &=& e^{-\frac{\pi \sqrt{3} }{8b} }.
\end{eqnarray}

Calculating also the prefactor, we  find
\begin{equation}
\langle I\rangle = \frac{\pi b}{5\sqrt{3}}  e^{-\frac{\pi \sqrt{3}} {8b}}.
\label{exp65}
\end{equation}

For $b\sim \beta/|k|^2$, we obtain the sub-critical exponents $\nu_{-}$ and $\mu_{-}$ listed in Eq.~(\ref{num-exp}). 
Finally, we use relations between parameters of the classical Hamiltonian and the driven LMGm from Eq.~(\ref{dictionary}) to find  
\begin{equation}
n_{\rm ex}=\frac{0.51\cdot S \beta}{B_x^{3/2}(B_x-D)^{1/2}}e^{-0.962\cdot (B_x-D)^2/\beta}.
\label{eq:nex_below}
\end{equation}

\subsection{The phase transition at $k>0$ ($D>B_x$ in LMGm)}

\begin{figure}[t!]
\includegraphics[width=3.0in]{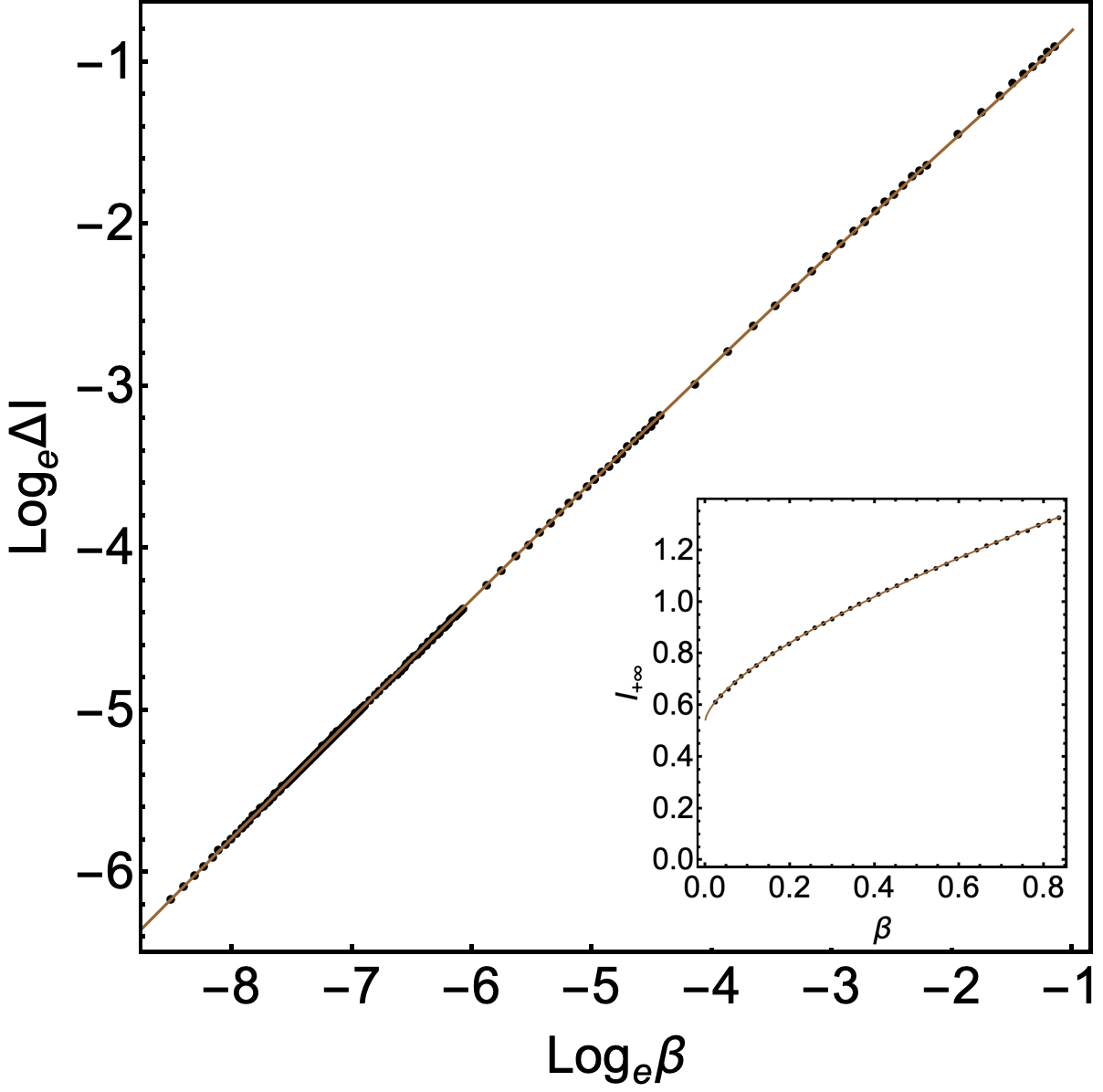}
\caption{Best fit of $\Delta I$, defined in Eq.~(\ref{I-uni}), in the log-log plot. The black points are the results of numerical simulations over a broad range of values of $\beta$. The brown curve is the best fit given by $\ln(\Delta I)=0.8\ln \beta+a(\ln \ln \beta)^r$, with $a=0.2$ and $r=1.5$, suggesting that $\nu_+=0.8$. The constant parameters were chosen so that $k=m=g=1$. The inset shows $I_{+\infty}$ versus $\beta \in (0.02,0.9)$. A projection of the best fit  to $\beta \rightarrow 0$ shows that $I_{+\infty}$ saturates at a finite value. Numerically obtained field theory curves for $I_{+\infty}$ are analogous to those in Fig.~\ref{fig:nex_vs_beta} (Right) for $n_{\rm ex}$ in LMGm.}
\label{power-log}
\end{figure}
 
In Eq.~(\ref{I-uni}) the second contribution is vanishing as $\beta \rightarrow 0$, but its prefactor depends on a smaller power of $k$. Note that at $\nu_+=3/4=0.75$, the coefficient at the $\beta$-dependent term in Eq.~(\ref{I-uni}) becomes $k$-independent, whereas the $\beta$-independent contribution vanishes as $k\rightarrow 0$. In Fig.~\ref{power-log}, we show results of our numerical simulations of the Newton's equation
\begin{equation}
\ddot{x}=bt+x-x^3,
\label{xt-eq}
\end{equation}
which is equivalent to the evolution equations (\ref{ham-em}). Having a time-dependent trajectory $x(t)$, we calculated the final action as an integral of $(\dot{x})^2$ over time along a periodic trajectory. We then extracted the $\beta$-dependent part of the adiabatic invariant, $\Delta I$, by fitting points at different $\beta$ using Eq.~(\ref{I-uni}).

In Fig.~\ref{power-log}, a correction $\Delta I$ is  shown in a log-log plot
with $\ln (\Delta I)$ and $\ln \beta$ on the axes. The results are fitted well with a power law
$\Delta I \approx c \beta^{0.73}$, suggesting that $\nu_+=0.73$. However, the coefficient $c$  showed an additional weak dependence on $\beta$. A three-parameter fit with a logarithmic correction, 
$\ln (\Delta I)=\nu_+ \ln\beta +a(\ln\ln(\beta))^r$, suggests that the true exponent is $\nu_+ =0.8$, but there is a  logarithmic correction to this scaling, with $a\approx 0.2$ and $r=1.5$.
In any case, the numerical results confirm that $\nu_+$ is close to the value $0.75$, at which the second term in Eq.~(\ref{I-uni}) does not depend on $k$.

\vspace{0.2cm}
\paragraph*{\bf Adiabatic limit.}
Basic evolution stages at $k>0$ are illustrated in Fig.~\ref{stages}(a-c).
For $b\rightarrow 0$, the system stays  in the local minimum of the potential
$$
V(x)=-btx-\frac{x^2}{2}+\frac{x^4}{4},
$$ 
until this minimum merges with a local maximum. At this moment, the second derivative of $V(x)$ with respect to $x$ is zero. Let $x=x_c$ be this point; then
$$
3x_c^2=1.
$$
There are two solutions to this equation. In our case, the system starts with $x<0$. Hence, for these initial conditions, we should select the negative solution:
$$
x_c=-1/\sqrt{3}.
$$
  The corresponding time moment, $t_c$, of crossing the critical point is determined by setting the first derivative of $V(x)$ to zero at $x=x_c$: 
\begin{equation}
b t_c= -x_c+x_c^3=\frac{2}{3\sqrt{3}}. 
\label{tc-eq}
\end{equation}
At $t_c$, we disregard the kinetic energy because the evolution has been quasi-adiabatic and the system started from the ground state. Thus, the energy at this moment is the potential energy:
$$
E=V(x_c)=\frac{1}{12}.
$$
This is also the energy of the escape trajectory, shown by dashed arrows in Fig.~\ref{stages}(c). One of the turning points of this trajectory is $x_c$, near which $E-V$ behaves like a cubic function: $E-V\propto (x-x_c)^3$. The other turning point is a simple zero of $E-V$, which is
$$
x_1=\sqrt{3}.
$$
Hence, 
$$
E-V(x)=-\frac{1}{4} (x-x_c)^3(x-x_1).
$$
Given the analogy with classical particle motion, the kinetic energy is ${p}^2/2$, and 
\begin{equation}
p(x)=\sqrt{2[E-V(x)]} = (x-x_c)\sqrt{(x-x_c)(x_1-x)/2}.
\label{px}
\end{equation}
The adiabatic invariant for this trajectory can be calculated explicitly:
\begin{equation}
     \mathcal{I}_0 =\frac{1}{2\pi} \oint p(x)\, dx= \frac{1}{\pi} \int_{x_c}^{x_1} dx \, p(x)=\left( \frac{2}{3} \right)^{3/2}. 
\label{dr-scale}
\end{equation}
This determines the numerical coefficient in Eq.~(\ref{I-uni}), so the constant part of $I_{+\infty}$ can be determined precisely.

\vspace{2mm}
\paragraph*{\bf The $\beta$-dependent contribution.}
Analytical estimates produce slightly different predictions for the power exponent $\nu_+$, depending on the relative strengths of the two terms in Eq.~(\ref{I-uni}).
First, consider the case of a relatively fast crossing rate, so that the constant contribution is negligibly small in comparison with purely dynamic effects. In this limit, we disregard the presence of the potential barrier and set $k=0$ in the original potential (\ref{V-lz}). The dynamics at such parameter values has been considered previously in \cite{SinitsynPokrovsky2026QuasiAdiabaticEffects}. At $k=0$, the variables can be rescaled so that the Hamiltonian does not have any free parameters. Therefore, the change in the adiabatic invariant can be found up to a numerical coefficient that was determined in \cite{SinitsynPokrovsky2026QuasiAdiabaticEffects}:
\begin{equation}
\Delta I  =0.646 \cdot\frac{\beta^{3/4} m^{7/8}}{g^{5/8}}.
\end{equation}
Thus, either for relatively large $\beta$ or at $k=0$, the scaling exponent is  $\nu_+=\nu=3/4$. The same exponent was found for the transition probability in the nonlinear Landau-Zener model in the critical regime \cite{Liu2002}, suggesting a similar field-theoretical interpretation of that result.

However, in the slowest regime, when $I_0\gg \Delta I$, the role of the potential barrier is essential. In this case, we first explore the dynamics near the critical time defined by Eq.~(\ref{tc-eq}) and shift the variables
$$
t\rightarrow t-t_c, \quad x\rightarrow x-x_c.
$$
Near $t=0$, a cusp appears then in the potential energy, which locally has a cubic shape:
\begin{equation}
\mathcal{H}(t)\approx \frac{p^2}{2}-b t x-\frac{x^3}{\sqrt{3}} .
\label{hgen-p1}
\end{equation}

Let us rescale the variables again:
\begin{eqnarray}
\label{rescalet}
 t&=&2^{1/5}3^{1/10} b^{-1/5}\tau, \\
 \label{rescale-2xp}
x&=&2^{2/5}3^{1/15} b^{2/5} X, \quad p=2^{2/5}3^{1/15}b^{3/5} P,
\end{eqnarray}
so that  the Hamiltonian  has only numerical coefficients: 
\begin{equation}
{\cal H}_{PI}(\tau)=\frac{P^2}{2}-\tau X-2X^3,
\label{hPI}
\end{equation}
and the corresponding Newton's equation of motion
\begin{equation}
\frac{d^2X}{d\tau^2}=\tau+6X^2,
\label{Painleve-I}
\end{equation}
is recognized as the famous Painlev\'e-I equation. It is known to be integrable in a certain sense \cite{Fokas2006}, but here we do not use this, as the physics outside the critical point also plays an important role.

For $\tau\rightarrow -\infty$, the minimum is changing  slowly:
$$
X=-\sqrt{|\tau|/6},
$$
so the nonadiabatic effects are negligible until the system reaches the vicinity of the critical point, near $\tau =0$. Since the Hamiltonian (\ref{hPI}) has only numerical coefficients, the change in the adiabatic invariant near the critical point in variables $(X,P)$ is $O(1)$. Returning to the original variables $(x,p)$, we find that the change in the adiabatic invariant upon reaching the critical point scales as
$$ 
\Delta I=\frac{1}{2\pi} \oint{p}\,dx \propto \beta^{1},
$$
which is different from and subdominant to the power-law scaling with an exponent between $0.7$ and $0.8$ that we found numerically. Thus, we conclude that the essentially nonadiabatic effects near the critical point make a negligible contribution to $\Delta I$ in the slow-transition limit.

Instead, to explain the smaller value of the numerically found exponent $\nu_+$, we note that Eq.~(\ref{rescalet}) defines a time scale for the transition through the critical point:  
\begin{equation}
\delta t_{tr}\propto b^{-1/5}.
\label{tpt}
\end{equation}
As $b$ decreases, this time diverges, i.e., the system spends an anomalously long time near the critical point. However, in our calculation of the coefficient $\mathcal{I}_0$ in Eqs.~(\ref{px}) and (\ref{dr-scale}), the potential $V(x,t)$ was assumed to be frozen at $t=t_c$. Instead, due to the delay in Eq.~(\ref{tpt}), the time-dependent part of the potential energy is modified by the time the system exits the critical region:
$$
\delta V(x) \approx -b x \delta  t_{tr} \propto b^{4/5}x. 
$$
In Eq.~(\ref{px}), this introduces a correction 
$p(x)\rightarrow p(x)+\delta p(x)$, where $\delta p(x) \propto \delta V(x)$, and when substituted to Eq.~(\ref{dr-scale}), this gives a correction 
$$
\mathcal{I}_0 \rightarrow \mathcal{I}_0+\Delta \mathcal{I},
$$
where $\Delta \mathcal{I} \propto b^{4/5}$, and hence
\begin{equation}
\Delta I \propto \Delta \mathcal{I} \propto b^{4/5} \propto \beta^{4/5},
    \label{deltaI-fin}
\end{equation}
which agrees with the exponent $\nu_+ \approx 0.8$ found in our numerical simulations. 

Summarizing, the power exponent $\nu_+$ emerges from two different effects that dominate depending on the relative values of $\Delta I$ and $I_0$. For relatively fast transitions, when $\Delta I\gg I_0$, we find $\nu_+=0.75$, and the $\beta$-dependent term arises from nonadiabatic effects analogous to those at second-order phase transitions. On the other hand, when $I_0$ dominates, the exponent becomes $\nu_+=0.8$ because critical slowing down leads to noticeable changes in the potential well around the new global minimum before the system falls into it.

Despite their different physical origins, the exponents $0.75$ and $0.8$ are close to one another. This is why a good fit to the numerical data for $\Delta I(\beta)$ can be obtained in Fig.~\ref{power-log} using a single power-law exponent supplemented by a logarithmic correction.

\section{Discussion}

Recently, Suzuki and Zurek extended the Kibble--Zurek phenomenology to first-order {\it thermal} phase transitions, with defect creation in agreement with Eq.~(\ref{rho-scale}) \cite{Suzuki-Zurek}. We showed that the Suzuki--Zurek scaling can be inferred from quantum-coherent field theories, explaining its relevance to quantum annealing computations.

Although  quantum annealing deals with a combinatorially complex phase space, our approach is universal and straightforward for both analytical and numerical studies. Our findings agree with D-Wave experiments that observed unsuppressed  contributions to the number of errors \cite{Bando2020UniversalityQuantumAnnealer}. Therefore, it is likely that, by attempting to detect the scaling of a second-order phase transition, these experiments observed  first-order critical effects. Their presence in D-Wave simulations is also supported by observations of magnetic hysteresis \cite{Pelofske2026Hysteresis}.

The first-order transition can occur when unintentional energy barriers are introduced by small parameter uncertainties or with increased system sizes \cite{Young2010FirstOrder}. Our theory predicts that unwanted first-order critical points can be avoided by suppressing these uncertainties below a certain critical value, at least in systems with infinite-range interactions. The power law at a fast sweep rate can then be transformed into beneficial exponential suppression of excitations in the quasi-adiabatic regime.

Many famous computationally hard problems can be mapped onto the problem of finding the ground state of an Ising spin Hamiltonian. Quantum annealing finds this ground state through a purely coherent quantum-mechanical process, which is described by field theory. Based on our findings, we speculate that the performance of algorithms more generally can be characterized by  universality classes of phase transitions, which are described by quantum field theories.

Namely,   field theories leading to different scalings of defects at quantum phase transitions should correspond, via the link to quantum annealing, to different convergence rates of computational algorithms. By developing this correspondence, we hope that the existence of new, faster algorithms can be inferred for some known computational problems.

\appendix
\begin{acknowledgements}
This work was carried out under the auspices of the U.S. DoE through the Los Alamos National Laboratory, operated by Triad National Security, LLC (Contract No. 892333218NCA000001). 
\end{acknowledgements}

\bibliography{ref2}

\end{document}